\newcommand{\peps}{{\epsilon}}

\newcommand{\hp}{\hat{\rm \bf n}}
\newcommand{\hO}{{\hat{\rm \bf e}_{z}\,\!}}
\newcommand{\hm}{{\hat{\rm \bf e}_{x}\,\!}}
\newcommand{\hn}{{\hat{\rm \bf e}_{y}\,\!}}
\newcommand{\kg}{{k_{\rm g}}}
\newcommand{\wg}{{\omega_{\rm g}}}
\newcommand{\wc}{{\omega_{\rm cut}}}
\newcommand{\mg}{{m_{\rm g}}}

\newcommand{\debug}\bf
\documentclass[onecolumn]{emulateapj}
\usepackage{graphicx}
\date{ }
\begin{document}
\title{Detecting massive gravitons using pulsar timing arrays}
\author{Kejia Lee \altaffilmark{1,2}, Fredrick A. Jenet \altaffilmark{3}, 
Richard H. Price \altaffilmark{3}, Norbert Wex \altaffilmark{2}, Michael Kramer 
\altaffilmark{1,2} }
\altaffiltext{1}{The University of Manchester, School of Physics and Astronomy, 
Jodrell Bank Centre for Astrophysics, Alan Turing Building, Manchester M13 9PL 
(kjlee@mpifr-bonn.mpg.de)}
\altaffiltext{2}{Max-Planck-Institut f\"ur Radioastronomie, Bonn 53121, Germany 
}
\altaffiltext{3}{Department of Physics \& Astonomy, University of Texas at 
Brownsivlle, Brownsville, TX
78520, United State}

\begin{abstract}

At the limit of weak static fields, general relativity becomes Newtonian
gravity with a potential field that falls off as inverse distance,
rather than a theory of Yukawa-type fields with finite range.
 General relativity also predicts
that the speed of disturbances of its waves is $c$, the vacuum light
speed, and is non-dispersive.

For these reasons, the graviton, the boson for general relativity,
can be considered to be massless. Massive gravitons, however, are features of 
some alternatives to general relativity. This has
motivated experiments and observations that, so far, have been consistent
with the zero mass graviton of general relativity, but further tests
will be valuable.

A basis for new tests may be the high sensitivity gravitational wave
experiments that are now being performed, and the higher sensitivity
experiments that are being planned. In these experiments it should be
feasible to detect low levels of dispersion due to nonzero graviton
mass. One of the most promising techniques for such a detection may be the
pulsar timing program that is sensitive to nano-Hertz gravitational
waves. 

Here we present some details of such a detection scheme. The pulsar
timing response to a gravitational wave background with the massive graviton
is calculated, and the algorithm to detect the massive graviton is
presented. We conclude that, with $90\%$ probability, massles
gravitons can be distinguished from gravitons heavier than $3\times
10^{-22}$\,eV (Compton wave length $\lambda_{\rm g}=4.1 \times 10^{12}$ km), if 
biweekly observation of 60 pulsars are performed for
5 years with pulsar RMS timing accuracy of 100\,ns. If 60 pulsars are
observed for 10 years with the same accuracy, the detectible graviton
mass is reduced to $5\times 10^{-23}$\,eV ($\lambda_{\rm g}=2.5 \times 10^{13}$ 
km); for 5-year observations of
100 or 300 pulsars, the sensitivity is respectively $2.5\times
10^{-22}$ ($\lambda_{\rm g}=5.0\times 10^{12}$ km) and $10^{-22}$ eV 
($\lambda_{\rm g}=1.2\times 10^{13}$ km).  Finally, a 10-year
observation of 300 pulsars with 100\,ns timing accuracy would probe
graviton masses down to $3\times 10^{-23}$\,eV ($\lambda_{\rm g}=4.1\times 
10^{13}$ km).
\end{abstract}
\maketitle

\section{Introduction}

Although a complete quantum version of gravitation has not yet been
achieved \citep{Smolin03, Kiefer06}, in the weak field, linearized
limit quantization of gravitation can be carried out. One can
therefore ask phenomenologically whether or not the graviton in a
gravity theory is massive \citep{RT08, GN08}. Up to now the most
successful theory of gravitation is general relativity \citep{Will93,
 Stairs03, Will06, Damour07, KW09}. The quantization of its weak
field limit (see \cite{Gupta52} and references therein) shows that its
gravitational interaction is mediated by spin-2 massless bosons, the
so called gravitons. Recently, attention has turned to this question
due to both observational and theoretical advances. 

On the theoretical side there is the issue of the vDVZ discontinuity
\citep{Iwasaki70,VdV70,Zakharov70}, the prediction that for a massive
graviton, light deflection is greater than in general relativity by the
factor 4/3, no matter how small the mass. If this were true, despite
its paradoxical nature, then classical tests of starlight deflection
would rule out massive gravitons. But recently, resolutions  have been
proposed to the vDVZ paradox, so that massive gravitons cannot be ruled
out \citep{Vainshtein72, Visser98, DDGV02, DKP03, FS02}.

On the theoretical side also, advances in higher-dimensional theories
have led to graviton mass-like features (e.g., DGP model by \cite{DGP00});
meanwhile, because of possible issues with standard gravitation from solar
system scale \citep{ALLLNT98,ACEEJ08} to cosmological scale \citep{SM02},
interest has been growing in alternative gravity theories, some of
which predict massive gravitons \citep{DGP00,AMd09, EPF10}. There is,
therefore, considerable motivation for experimental or observational
tests of whether the graviton can be massive.

The literature contains many estimates for an upper limit on the graviton
mass, estimates that differ by many orders of magnitude.  An early limit
of $8\times 10^{4}$ eV \citep{Hare73} was based on considerations of a
graviton decay to two photons. At about the same time \cite{GN74} used the
assumption that clusters of galaxies are bound by more-or-less standard
gravity to infer an upper limit of $2\times 10^{-29} h_{0}$ eV, where
$h_0$ is the Hubble constant in units of 100\,km\,s$^{-1}$\,Mpc$^{-1}$. By
using the effect of graviton mass on the generation of gravitational waves
by a binary, and the rate of binary inspiral inferred from the timing
of binary pulsars, \cite{FS02} inferred an upper limit of $7.6\times
10^{-20}$ eV.  \cite{CJMM04} considered the effect of a graviton mass on
the power spectrum of weak lensing; with assumptions about dark energy
and other paramters, they estimated a  upper limit of $7\times 10^{-32}$
eV for the graviton mass. Reviews about these techniques can be found,
e.g.\,, in \cite{GN08, PDG08, Will06}.

These upper limits are all of value, since they are based on different
assumptions about the phenomenological effects of graviton mass. Most
of the estimates are based on the effects of graviton mass on the static
field of a source, typically assuming a Yukawa potential, though this may
not be the general case for a particular theory of gravity, such as fifth
force theories, or MOND-type theories \citep{Will98,Maggiore08}. In view
of the lack of a theory of the graviton, it is important to have upper
limits based on different phenomenological implications of graviton mass.

The mass limit of \cite{FS02} is based on the effect of graviton mass on
the generation of gravitational waves, not on their propagation, but the
dispersion relation for propagation is also an important independent approach
to a mass limit, as has been recently suggested by a number of groups
\citep{Will98, LH00, CHL03, SW09}. Questions about this method are timely
since the detection of gravitational waves is expected in the near future,
thanks to the progress with present ground-based laser interferometers,
possible future space-based interferometers \citep{HR00, JRS05}, and
pulsar timing array projects \citep{SBF93, SKLDJ06, Manchester06, HBBC09}.

The pulsar timing array is a unique technique to detect nano-Hertz
gravitational waves by timing millisecond pulsars, which are very
stable celestial clocks. It turns out that a stochastic gravitational
wave background leaves an angular dependent correlation in pulsar
timing residuals for widely spaced pulsars \citep{HD83, LJR08}. That
is, the correlation $C(\theta)$ between timing residual of pulsar
pairs is a function of angular separation $\theta$ between the
pulsars. One can analyse the timing residual and test such a
correlation between pulsar timing residuals to detect gravitational
waves \citep{JHLM05}. We find in this paper that if the graviton mass
is not zero, the form of $C(\theta)$ is very different from that
given by general relativity. Thus by
measuring this graviton mass dependent correlation function, we can
also detect the massive graviton.

The outline of this paper is as follows. The mass of the
graviton is related to the dispersion of gravitational waves in
\S~\ref{sec:gwmas}. The pulsar timing responses to a plane gravitational
wave and to a stochastic gravitational wave background in the case
of a massive graviton are calculated in \S~\ref{sec:sigres}. The
massive graviton induces effects on the shape of pulsar timing
correlation function, which is derived in \S~\ref{sec:hel}, while the
detectability of a massive gravitational wave background is studied in
\S~\ref{sec:sigdec}. The algorithm for detect massive graviton using a
pulsar timing array, and the sensitivity of that algorithm are examined
in \S~\ref{sec:memg}. We discuss several related issues, and conclude
in \S~\ref{sec:conc}.

\section{Gravitational Waves with Massive Gravitons} \label{sec:gwmas}

We incorporate the massive graviton into the linearized weak field
theory of general relativity \citep{Gupta52,AD59, Weinberg72}. For
linearized gravitational waves, specifying the graviton
mass is equivalent to specifying the gravitational wave (GW)
dispersion relation that follows from the special relativistic relationship
\begin{equation}
	E^2=p^2 c^2+m^2 c^4, \label{eq:emrel}
\end{equation} 
where $c$ is the light velocity, $E$ is energy of the particle, and
$p$ and $m$ are the particle's momentum and rest mass
respectively. One can derive the corresponding dispersion relation
from Eq.~(\ref{eq:emrel}) by replacing the momentum by ${\bf p}=\hbar
{\bf k}_g $ and the energy by $E=\hbar \omega_g $, where $\hbar$ is
the reduced Planck constant, with ${\bf k}_g$ and $\omega_g$
respectively the GW wave vector and the angular frequency. With these
replacements, the dispersion relation for a massive vacuum GW graviton
propagating in the $z$ direction reads
\begin{equation}
 {\bf \kg}(\wg) = \frac{\left(\wg^{2}-\wc^2\right)^{\frac{1}{2}}}{c}\, \hO\,,
	\label{eq:dispf}
\end{equation}
where $\hO$ is the unit vector in the $z$ direction. If the
gravitational wave frequency $\wg$ is less than the cut-off frequency
$\omega_{\rm cut}\equiv\mg c^2/\hbar$, then the wave vector becomes
imaginary, indicating that the wave attenuates, and does not
propagate. (The equivalent phenomena for electromagnetic waves can be
found in \S 87 of \cite{LandauEM}).

At a space-time point $(t, {\bf r})$, the spatial metric perturbation due 
to a monochromatic GW is \begin{equation}
 h_{ab}(t,{\bf r})=\Re\left[\sum_{P=+,\times} A^{P} \peps^{P}_{ab} e^{i[\wg 
 t-{\bf r} \cdot {\bf \kg}(\wg) ] }\right],\label{eq:planwv}
\end{equation}
where $\Re$ indicates the real part, and where the $a, b$ range over
spacetime indicts from 0 to 3. The summation is performed over the
polarizations of the GW. Since we are not assuming that
 general relativity is the theory of gravitation, we could, in
 principle, have as many as six polarization states. For
 definiteness, however, and to most clearly show how pulsar timing
 probes graviton mass, we will confine ourselves in this paper to
 only the two standard polarization modes of general relativity, denoted
$+$ and $\times$, the usual `TT'' gauge (see Appendix A \ref{app:sing} for the 
details). Thus, the polarization index takes on only the values
$P=+, \times$, with $A^{P}$ and $\peps^{P}$ standing for the
amplitude and polarization tensors for the two transverse traceless
modes.

The polarization tensor $\peps^{P}$ is described in terms of an
orthonormal 3-dimensional frame associated with the GW propagating
direction. Let the unit vector in the direction of GW propagation be $\hO$; we 
can choose the other two mutually orthogonal unit vectors $\hm, \hn$
to be both perpendicular to $\hO$.
In terms of these three vector, $\hO, \hm$ and $\hn$, the polarization
tensors are given as 
\begin{eqnarray} \peps^{+}_{ab}=\hm_{a} \hm_{b}
 -\hn_{a}\hn_{b}, \nonumber\\ \peps^{+}_{ab}=\hm_{a} \hn_{b}
 +\hn_{a}\hm_{b}\label{eq:polten}\,.
	\end{eqnarray}

Since the polarization tensors are purely spatial, we will have only
spatial components of the metric perturbations. For a stochastic
gravitational wave background, these metric perturbations are a
superposition of monochromatic GWs with random phase and amplitude, and can be 
written as
\begin{equation}
  h_{ij}(t,r^i)=\sum_{P=+,\times}\int_{-\infty}^{\infty}df_{\rm g}\int d\Omega\, 
  h^{P}(f_{\rm g},\hO )\, \peps^{P}_{ij}(\hO) e^{ i[\wg t-{\bf 
  k}_g(\wg)\cdot{\bf r }} ],
	\label{eq:grbk}
\end{equation}
where $f_{\rm g}={\wg}/{2\pi}$ is the GW frequency, $\Omega$ is solid
angle, spatial indices $i,j$ run from 1 to 3, and $h^{P}$ is the amplitude of the
gravitational wave propagating in the direction of $\hO$ per unit
solid angle, per unit frequency interval, in polarization state $P$. If the
gravitational wave background is isotropic, stationary and independently
polarized, we can define the characteristic strain $h_{c}^{P}$
according to \citep{Ma00, LJR08}, and can write
\begin{equation}
  \langle h^{P}(f_{\rm g},\hO) h^{\star P'}(f_{\rm g}', \hO')\rangle= 
  \frac{|h_{c}^{P}|^2}{16 \pi f_{\rm g}} \delta_{PP'} \delta(f_{\rm g}-f_{\rm 
  g}')\delta(\hO-\hO'), \label{eq:hcdef}
\end{equation}
where the $\star$ stands for the complex conjugate and $\langle
\rangle$ is the statistical ensemble average. The symbol $\delta_{PP'}$
is the Kronecker delta for polarization states; $\delta_{PP'}=0$ when
$P$ and $P'$ are different, and $\delta_{PP'}=1$, when $P$ and $P'$
are the same. With the relationships above one can show that
\begin{equation}
	\langle 
	h_{ab}(t)h_{ab}(t)\rangle=\sum_{P=+,\times}\int_{0}^{\infty}\frac{|h_{c}^{P}|^{2}}{f_{\rm 
	g}}
	df_{\rm g}\,.
\end{equation}

\section{Single Pulsar Timing Residuals Induced by a Massive Gravitational Wave 
Background}

\label{sec:sigres}

We now turn to the calculation of the pulsar timing effects due to the stochastic 
gravitational wave background prescribed by Eq.~(\ref{eq:grbk}). That
background is composed of monochromatic plane wave with random phase and with
amplitude determined from some chosen spectrum. We can then first calculate 
the pulsar timing effect due to a monochromatic plane wave, and then 
add together all contributions to find the effects of a stochastic background.

In the weak field case, the time component of the null geodesic equations for 
photons
is \citep{Liang00a,HJL09}

\begin{equation}
	\frac{d \omega}{d \lambda}+\frac{\omega^2}{2} \frac{\partial{h_{ij}} 
	}{\partial t}
	\hp^{i}\hp^{j}=0\,.
\end{equation}
Here
 $\omega$ is the angular frequency for the photon, $\hp^{i}$ is the 
direction from the observer (Earth) to the photon source (pulsar),
and $\lambda$ is an affine parameter along the photon trajectory, normalized
so that in the Minkowski background $dt/d\lambda =\omega$. From this 
geodesic equation, the frequency shift of a pulsar timing signal induced by a 
monochromatic, plane, massive graviton GW is 
\begin{equation}\label{delomega}
\frac{\Delta\omega(t)}{\omega}=-\,\frac{ \hp^{i}\hp^{j} }{2\left 
 (1+({c}/{\wg}){\rm \bf \kg} \cdot \hp\right )} \left[ h_{ij}(t,0)-h_{ij}(t-
 |{\rm \bf D}|/c,{\rm \bf D} )\right],\label{eq:z}\end{equation}
where it is assumed that the observer is at the coordinate origin and that $\rm \bf 
D$ is the displacement vector, in the background, from the observer to the pulsar. 
Equation~(\ref{delomega}) is a minor generalization of the relationship for 
zero-mass gravitons given in many references 
(\cite{EW75,Sazhin78,Detweiler79,LJR08}).
It is clear that the 
frequency shift of the pulsar timing signal only involves the metric perturbations at the 
observer, i.e., the term $h_{ij}(t,0)$, and the metric perturbation at the 
pulsar, i.e., the term $h_{ij}(t,\rm \bf D)$. The dispersion relation of the GWs
enters, 
through the denominator, in the geometric factor $(1+c {\rm \bf\kg}\cdot 
\hp/\wg )$ and well as the phase difference between the GW at the earth 
$h_{ij}(t,0)$ and the GW at the pulsar $h_{ij}(t-
|{\rm \bf D}|/c,{\rm \bf D} )$. The induced pulsar
timing residuals R(t) are given by the temporal integration of the
above frequency shift at Earth, thus 
\begin{equation}
	R(t)=\int_{0}^{t} \frac{\Delta \omega(\tau)}{\omega} d\tau.
	\label{eq:r}
\end{equation}

The equations Eq.~(\ref{eq:grbk}), (\ref{eq:z}), and (\ref{eq:r}) determine the 
response of pulsar timing to a monochromatic plane GW. One can show the 
pulsar timing residual $R(t)$ induced by the stochastic GW background is

\begin{equation}
  R(t)=\frac{i}{2}\sum_{P=+,\times}\int_{-\infty}^{\infty} df_{\rm g} \int 
  d\Omega \frac{ \hp^{i}\hp^{j} }{\left (\wg+c {\rm \bf \kg} \cdot \hp\right )} 
  h^{P}(f_{\rm g},\hO)\peps^{P}_{ij}(\hO) \left[1- e^{-i {\rm \bf \kg(\wg) \cdot 
  D}}\right]\left[1-e^{i\wg t}\right] \label{eq:rbk}.
\end{equation}
This is, for example, a minor modification of Eq.~(A9) given by \cite{LJR08}.  
{ As we pointed out shortly before, the massive graviton dispersion relation 
enters via the term $\wg+c {\rm \bf \kg} \cdot \hp$ as in Eq.(~\ref{eq:z}) and 
the term  $1- e^{-i {\rm \bf \kg(\wg) \cdot D}}$, which comes from the phase 
difference between pulsar term and earth term.}

After the nonobservable zero frequency component is removed, the
auto-correlation function 
$\langle 
R(t)R(t+\tau)\rangle$ can be calculated by
replacing the statistical ensemble average with the time average,
\begin{equation}
	\langle R(t)R(t+\tau)\rangle = \sum_{P=+,\times} \int_{-\infty}^{\infty} 
	df_{\rm g} \int d\Omega\, \frac{|h_{c}^{P}|^2}{32 \pi f_{\rm g} } 
  \left[\frac{\peps ^{P}_{ij}\hp^i \hp^j }{\wg+c {\rm \bf \kg \cdot 
  \hp}}\right]^2 \left[1-\cos\left(\frac{D}{c} (\wg+c{\rm \bf \kg \cdot 
  \hp})\right) \right] e^{i\wg \tau}
	\label{eq:autocor}\,,
\end{equation}
where $D=|{\rm \bf D}|$ is the pulsar distance.
The folded timing residual power spectra $S_{\rm R}(f_{\rm g} )$ is defined to 
be the
Fourier transform of the autocorrelation function of the timing residual, 
\begin{equation}
  S_{\rm R}(f_{\rm g} )=2\int _{-\infty}^{\infty} \langle R(t)R(t+\tau)\rangle 
  e^{-2\pi i f_{\rm g} \tau } d\tau, \end{equation}
for which the explicit result is
\begin{equation}
  S_{\rm R}(f_{\rm g} )=\sum_{P=+,\times} \int d\Omega\, 
  \frac{|h_{c}^{P}|^2}{16 \pi f_{\rm g} } \left[\frac{\peps ^{P}_{ij}\hp^i 
  \hp^j }{\wg+c {\rm \bf \kg \cdot \hp}}\right]^2 \left[1-\cos\left(\frac{D}{c} 
  (\wg+c{\rm \bf \kg \cdot \hp})\right) \right]. \label{eq:powr}\end{equation}

For a power-law stochastic GW background, the power spectra of the induced 
pulsar timing residual is (see Appendix for the details)
\begin{equation}
  S_{\rm R}(f_{\rm g})=\sum_{P=+,\times}\frac{|h_{c}^{P}(f_{\rm g})|^2}{24 \pi^2
  f_{\rm g}^3} \eta(f_{\rm g}),
\label{srfg}
\end{equation}
where the $\eta(f_{\rm g})$ is
\begin{displaymath}
  \eta(f_{\rm g})=\left\{\begin{array}{c c}\frac{-4 \zeta ^3+6 \zeta +3 
	\left(\zeta ^2-1\right) \log \left(\frac{\zeta +1}{1-\zeta }\right)}{2 
	\zeta^5} & \textrm{if $f_{\rm g}> f_{\rm cut}$}
		\\
		0 &\textrm{if $f_{\rm g}\le f_{\rm cut}$}
	\end{array} \right.
\end{displaymath}
and \begin{equation}
  \zeta =\sqrt{1-\frac{\omega_{\rm cut}^2}{\wg^2}}\quad\quad\quad \omega_{\rm 
  cut}\equiv\mg c^2/\hbar\,.
\end{equation}

\begin{figure}[ht]
	\centering \includegraphics[totalheight=2.5in]{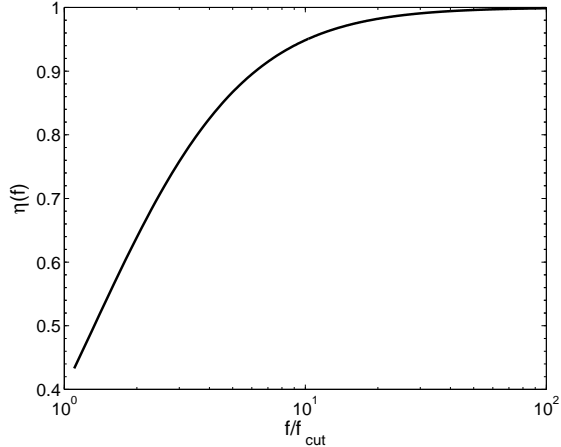}
  \caption{The value of $\eta(f)$ as a function of GW frequency
   $f$ in units of the cut-off frequency $f_{\rm cut}$. When the
   GW frequency becomes much larger than the cut-off frequency,
   $\eta$ approaches unity, which means that the timing
   residuals approach those of the massless GW case.  At the other extreme, for 
   $f_{\rm g}$ close to $f_{\rm cut}$, the power in the timing residuals is 
   minimal.}
		\label{fig:eta}
\end{figure}

One can see that $\eta(f_{\rm g})$ is the spectral correction factor for massive 
GW backgrounds, if compared with the case of general relativistic GW background 
\citep{LJR08}, where $\eta$ is equal to unity. To show the frequency dependence
of the graviton mass effect, $\eta(f_{\rm g})$ is plotted in Fig.~\ref{fig:eta}.

When $f_{\rm g}>f_{\rm cut}$, a convenient polynomial approximation of $\eta(f)$ 
with maximal error of
$1.5\%$ is \begin{equation}
  \eta(f_{\rm g})=-0.7764 \left(\frac{f_{\rm cut}}{f_{\rm g}}\right)^4+1.748 
  \left(\frac{f_{\rm cut}}{f_{\rm g}}\right)^3-1.001
  \left(\frac{f_{\rm cut}}{f_{\rm g}}\right)^2-0.5868\, \frac{f_{\rm 
  cut}}{f_{\rm g}}+1.012\,.
	 \label{eq:etaapp}
\end{equation}

{ A least-squares polynomial fitting technique was used to calculate the 
coefficients in the above equation. } The RMS level $\sigma_{\rm R}$ of the 
timing residual power is defined
to be $\sigma_{\rm R}^2 =\int_{0}^{\infty} S_{\rm R}(f)
df$. The spectra of GW backgrounds generated by various astrophysical
processes are usually summarized as power-law spectra with power index
$\alpha$, i.e. the characteristic strain of GWs is 
$h_{\rm c}=A_{\rm c} (f/f_{0})^{\alpha}$. For such power-law spectra,
the RMS level of corresponding pulsar timing residuals is

\begin{equation}
	\sigma_{\rm R}^2\simeq\frac{A_{\rm c}^2 f_{\rm L}^{2\alpha}}{f_{0}^{2 
	\alpha}}\left(
	\beta_1 \frac{f_{\rm cut}^4}{f_{\rm L}^{6}}
	+\beta_2 \frac{f_{\rm cut}^3}{f_{\rm L}^{5}}
	+\beta_3 \frac{f_{\rm cut}^2}{f_{\rm L}^{4}}
	+\beta_4 \frac{f_{\rm cut}}{f_{\rm L}^{3}}
	+\beta_5 \frac{1}{f_{\rm L}^{2}}
	\right),
	\label{eq:sign}
\end{equation}
where the constants $\beta_1 \dots \beta_{5}$ take following values
\begin{eqnarray}
	\beta_1 &=&3.278 \times 10^{-3} (\alpha-3)^{-1}, \nonumber \\
	\beta_2 &=&1.476 \times 10^{-2} (5-2\alpha)^{-1}, \nonumber \\
	\beta_3 &=&4.226 \times 10^{-3} (\alpha-2)^{-1}, \nonumber \\
	\beta_4 &=&4.955 \times 10^{-3} (2\alpha-3)^{-1}, \nonumber \\
	\beta_5 &=&4.272 \times 10^{-3} (1-\alpha)^{-1}, \end{eqnarray}
and the $f_{\rm L}={\rm Max}[T^{-1},f_{\rm cut}]$ is the larger of the following
two frequencies: (1)~the frequency cut-off ($T^{-1}$) due to the finite length time 
span $T$ of observation; (2) ~the intrinsic frequency cut-off $f_{\rm 
cut}=\wc/(2\pi)$ due to graviton mass. 

One can derive an upper limit for the GW velocity using single pulsar
timing data, because of the surfing effect \citep{BPPP08}; but it is
unlikely that one can use single pulsar timing data to constrain the
graviton mass \citep{BPPP08b}. Because of the correction factor
$\eta(f_{\rm g})$ (see Fig.~\ref{fig:eta}), the graviton mass reduces the GW 
induced
pulsar timing residuals. This prevents us from constraining the
graviton mass using the amplitude of single pulsar timing
residuals. However, as explained in the next section, the cross
correlation between pulsar timing residuals from different directions
will help us in detecting the graviton mass.

\section{The Angular Dependent Correlation Between Pulsars}

\label{sec:hel}

A stochastic GW background leaves a correlation between timing residuals
of pulsars pairs \citep{HD83, LJR08}. Such a correlation,
$C(\theta)$, depends on the angular distance $\theta$ between two pulsars. It 
turns out that the graviton mass changes the shape
of this correlation function. One can therefore detect a massive
graviton by examining the shapes of pulsar timing correlation functions.

\begin{figure*}[ht]
	\centering
		\includegraphics[totalheight=2.5in]{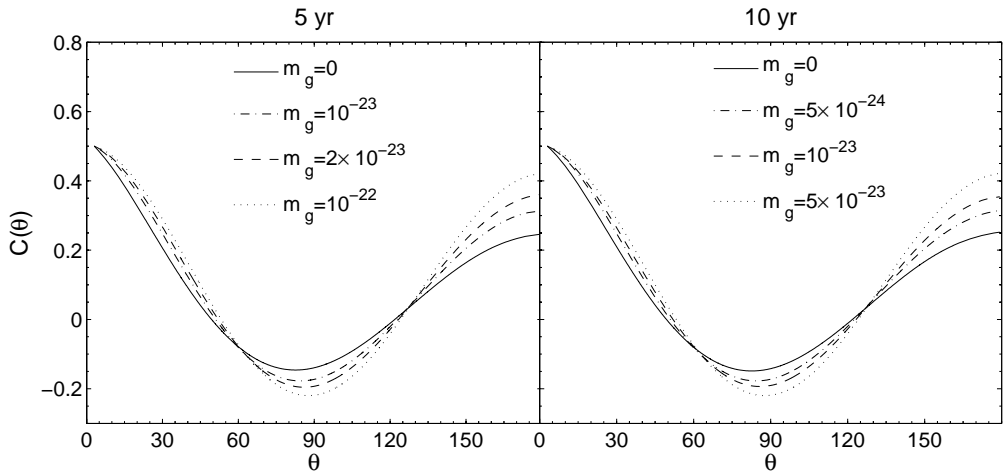}
	\caption{The atlas for cross-correlation functions
	$C(\theta)$. The label of each curve indicates the corresponding
	graviton mass in units of electron-volts (eV). The left panel shows
	the correlation functions for a 5 year bi-weekly observation.
	The right panel shows correlation functions for 10-years of bi-weekly
	observations. We take $\alpha=-2/3$ for these results. These correlation are 
	normalized such that the $C(0)=0.5$
	for two \emph{different} pulsars.} \label{fig:hel}
\end{figure*}

As shown in Appendix~\ref{app:hel}, the pulsar timing
cross-correlation function for a massive GW background depends on the
graviton mass, specific power spectra of GW background, and
observation schedule. In this way, an analytical expression for the
cross-correlation function would not be possible. We use Monte-Carlo
simulations in this paper to determine the shape of the correlation
function for GW backgrounds with a power-law spectra. In the Monte-Carlo
simulations for $C(\theta)$, we randomly choose pulsars from an
isotropic distribution over sky positions. We then hold constant these
pulsar positions and calculate the angular separation $\theta$ between
every pair of pulsars. Next, to simulate the power-law GW background, we
generate $10^{4}$ monochromatic waves, choosing random phase, and choosing
the amplitude from the power-law $h_{\rm c}=A_{\rm c} (f/f_{0})^{\alpha}$, 
where we take
$\alpha=-2/3$ \citep{Phinney01,JB03, WL03, EINS04, SHMV04, WLH09}.
The timing residuals are calculated using Eq.~(\ref{eq:z}) and (\ref{eq:r}). 
Then, the
cross-correlation function, $C(\theta)$, between pulsar pairs is calculated. We
repeat such processes and average over the angular dependent
correlation function $C(\theta)$ until the change in $C(\theta)$ is
less than $0.1\%$. The averaged correlation function is then smoothed
by fitting an eighth order Legendre polynomial (see \cite{LJR08} for the
details). This smoothed $C(\theta)$ is the correlation function we
need. The $C(\theta)$ are plotted for various parameters in
Fig.~\ref{fig:hel}. We also check the results by choosing different
sets of pulsars to make sure the $C(\theta)$ is not sensitive to the
details of the random pulsar samples. As the $C(\theta)$ is the
statistical expectation of the correlation function, we call it the
\emph{theoretical correlation function} in contrast with the
\emph{observed correlation function} defined in the next section.

As one may expect, the massive graviton has stronger effects for data from long 
observing periods than from short periods. One can see this by comparing the 
5-year and 10-year correlation functions given in Fig.~\ref{fig:hel}, where 
curves corresponding to the same range of graviton mass show considerably
greater deviations from the massless case in the 10-year correlation 
function than in the 5-year one. 

\section{An Estimate of the Detectability of a Gravitational Wave Background 
with a Massive Graviton}
\label{sec:sigdec}

As we have explained, the massive graviton reduces the pulsar timing response to 
GWs through the correction factor $\eta(f_{\rm g})$. In this way, a non-zero 
graviton mass reduces the pulsar timing array sensitivity for detecting a GW 
background.  It is interesting to know how the sensitivity changes, if the 
graviton is massive.

The stochastic GW background is detected by comparing the measured 
cross-correlation function $c(\theta_{l})$ with the theoretical correlation 
$C(\theta)$ calculated in previous section. Here, the measured cross-correlation 
function $c(\theta_{l})$ is defined by
\begin{equation}
	c(\theta_{m})=\frac{
	\sum_{l=0}^{N-1}
	(R_{a}(t_{l})-\overline{R_{a}(t_{l})}\;)(R_{b}(t_{l})-
\overline{R_{b}(t_{l})}\;)}{\sqrt{\sum_{l=0}^{N-1}(R_{a}(t_{l})-
\overline{R_{a}(t_{l})}\;)^{2}\sum_{l=0}^{N-1}(R_{b}(t_{l})-\overline{R_{b}(t_{l})}\;)^{2}}}\ ,
	\label{eq:obsc}
\end{equation}
where the $R_{a}(t_{l})$ and $R_{b}(t_{l})$ are the timing residuals of
pulsar `a' and `b' at time $t_{l}$ and $N$ is the number of
observations. The $\overline{R_{b}(t_{l})}=\sum_{l=0}^{N-1} R_{b}(t_{l})/N$ and
$\overline{R_{a}(t_{l})}=\sum_{l=0}^{N-1} R_{a}(t_{l})/N$.
The $\theta_{m}$ is the angle between the direction
pointing to pulsar `a' and the direction pointing to pulsar
`b'. Given $N_{\rm p}$ pulsars, the index $m$ runs from 1 to the number of 
pulsar pairs
$M=(N_{\rm p}-1)N_{\rm p}/2$, because the autocorrelations are not
used. 

Following \cite{JHLM05}, we define \begin{equation}
 \rho=
 \frac{
	\sum_{m=1}^{M}\left(C(\theta_{m})-\overline{C}\right)\left(c(\theta_{m})-\overline{c}\right)}{\sqrt{\sum_{m=1}^M 
	\left(C(\theta_{m})-\overline{C}\right)^2\sum_{m=1}^M\left(c(\theta_{m})-\overline{c}\right)^2}}\,,
\label{eq:s}
\end{equation}
where $\overline{C}=\sum_{m=1}^{M}C(\theta_{m})/M$ and
$\overline{c}=\sum_{m=1}^{M}c(\theta_{m})/M$. Then the statistic $S$,
describing the significance of the detection, is $S=\sqrt{M\;}\rho$. In 
particular, when there is no GW present, the $c(\theta_m)$ will be 
Gaussian-like white noise, the probability of getting a detection significance 
larger than $S$ is about ${\rm erfc}(S/\sqrt{2})/2$ \citep{JHLM05}.

Our aim is to determine the ability of a given 
pulsar timing array configuration
to detect a GW background.
 To do this, we calculate the expected value 
for the detection significance $S$ by using a second set of Monte-Carlo 
simulations. These second Monte-Carlo simulations are similar to the first ones, 
but instead of calculating the average value for $C(\theta)$, we inject white 
noise for each pulsar, to represent the intrinsic pulsar noise and 
instrumental noise, and we calculate the expected value of 
$S$. We summarize the steps here.

1. Generate a large number of GW source ($10^4$) to simulate the required GW 
background.

2. Calculate the timing residual for each pulsar as described above, and add white 
Gaussian noise.

3. Calculate the measured correlation $c(\theta_{m})$ using Eq.~(\ref{eq:obsc}), 
and calculate the detection significance $S$ using Eq.~(\ref{eq:s}).

4. Repeat steps 1,2,3 and average over the detection significance $S$. The converged 
$S$ is the value needed to estimate the detection significance. 

\begin{figure*}[ht]
	\centering
		\includegraphics[totalheight=2.5in]{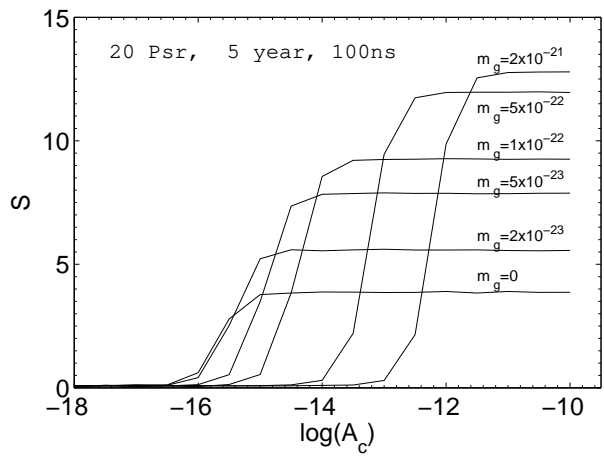}
    \caption{The expected GW background detection
	 significance using a pulsar timing array with 20
     pulsars, observed for 5 years, with 100\,ns timing
     noise. The graviton mass, in units of electron
     volts, is labeled above each curve. The $x$-axis is
     the amplitude for the characteristic strain of the
	 GW background ($f_{0}=1\, {\rm yr}^{-1}$,
     $\alpha=-2/3$), while the $y$-axis is the expected
     detection significance $S$.}
	\label{fig:avs}
\end{figure*}

The results for the expectation value of $S$, as a function of GW
amplitude $A_{\rm c}$ for various pulsar timing array configurations,
are presented in Fig.~\ref{fig:avs}. We have also compared
simulations from several different pulsar samples with the same number
of pulsars to make sure such $S$ is not sensitive to the detailed
configuration of the pulsar samples.

Two features of the curves in Fig.~\ref{fig:avs} are worth
noting. First, the minimal detection amplitude of a GW background
becomes larger, when a massive graviton is present, i.e., the leading
edge of the $S$-$A_{\rm c}$ curve shifts rightwards as $m_{\rm g}$ is
made larger. This tells us that in order to detect a massive GW
background one needs a stronger GW background signal or a smaller
pulsar intrinsic noise than in the case of a massless GW background. As 
previously
noted, this effect is mainly due to the reduction of the pulsar timing
response and the reduction of the gravitational wave amplitude at
lower frequencies. Fig.~\ref{fig:avs} also tells us when we can
neglect the effect of a massive graviton. It is clear from
Fig.~\ref{fig:avs} that if $m_{\rm g}\le 2\times 10^{-23}$ eV for a
5-year observation, the minimal detection amplitude is not reduced by more
than 5\%. For 10 years of observation a 5\% reduction corresponds to
$m_{\rm g}=10^{-23}$ eV.

The second noteworthy feature of the $S$-$A_{\rm c}$ curves in
Fig.~\ref{fig:avs} is that of the saturation level of detection
significance. Due to the pulsar distance term of Eq.~(\ref{eq:rbk})
(the term involving the $\rm \bf D$), the detection significance
achieves a saturation level when the GW induced timing residuals are
much stronger than the intrinsic pulsar timing noise \citep{JHLM05}. From
Fig.~\ref{fig:avs}, we note that the saturation level of detection
significance is large, when the graviton is massive, i.e., the plateau
at the right part of the $S$-$A_{\rm c}$ curve becomes higher for a
more massive graviton. This is rather similar to the whitening filter
discussed by \citep{JHLM05}. The graviton mass introduces a low
frequency spectral cut-off, which is equivalent to applying a
whitening filter to the timing signals. In this way, the saturation
level of $S$ starts to grow for a massive GW background. Because the
cut-off in the frequency domain coherently removes low frequency GW
components, the envelope of the these $S$-$A_{\rm c}$ curves is
similar to the curves with the whitening filter as described by
\cite{JHLM05}.

\section{Estimate of the Detectability of the Massive Graviton}
\label{sec:memg}

In \S\,\ref{sec:sigdec}, we have discussed how the detection significance for 
the stochastic GW background is affected by the graviton mass. { Here we 
formulate the qustion as a detection problem rather than a parameter estimation 
problem.  In this section, we will construct a detector, which accepts pulsar 
timing array data and then determines if the gravitational wave background is 
massless or massive. Then we simulate pulsar timing data, passing through the 
detector, to determine the quality of such detector. With the detector quality, 
we then discusse related technical requirements (the number of pulsars, the 
pulsar noise level, and so on) for performing such detections.}

The question, whether the graviton mass is zero or not, is best formulated as a 
statistical hypothesis test composed of two hypotheses F and H given by 

\begin{displaymath}
	\left\{ \begin{array}{c}
		\textrm{H: The graviton is massless;} \\
		\textrm{F: The graviton is massive.}
\end{array} \right.
\end{displaymath}

The detection algorithm determines which hypothesis is accepted. We
define the detection rate $P_{\rm d}$ as the probability that the
detection algorithm gives statement F, when the graviton is massive,
i.e., $P_{\rm d}=P(F|m_{\rm g}>0)$; and we define the false alarm rate
$P_{\rm f}$ as the probability of getting statement F when the
graviton is massless, i.e., $P_{\rm f}=P(F|m_{\rm g}=0)$. The quality
of the detection algorithm is evaluated by calculating the relation
between the detection rate $P_{\rm d}$ and the false alarm rate
$P_{\rm f}$. Here we take the standard approach \citep{DR68} that one
fixes $P_{\rm f}$ to a certain threshold level $P_{\rm th}$ and
calculates the corresponding $P_{\rm d}$. Throughout this paper we fix
$P_{\rm th}=0.1\%$. If a detector maximizes the $P_{\rm d}$ for a
prescribed value of $P_{\rm f}$, we say that such detector is optimal
\citep{Kassam88}.

As we have shown in \S\,\ref{sec:hel}, the graviton mass changes the shape of 
the correlation function. The best way to detect such a difference is to use 
following statistics \citep{Kassam88} \begin{equation}
	\gamma=\frac{
	\sum_{m=1}^{M}\left(C_{\rm m}(\theta_{m})-\overline{C_{\rm 
	m}}\right)\left(c(\theta_{m})-\overline{c}\right)}{\sqrt{\sum_{m=1}^M 
	\left(C_{\rm m}(\theta_{m})-\overline{C_{\rm 
	m}}\right)^2\sum_{j=1}^M\left(c(\theta_{m})-\overline{c}\right)^2}}-\frac{
	\sum_{m=1}^{M}\left(C_0(\theta_{m})-\overline{C_0}\right)\left(c(\theta_{m})-\overline{c}\right)}{\sqrt{\sum_{j=1}^M 
	\left(C_0(\theta_{m})-\overline{C_0}\right)^2\sum_{j=1}^M\left(c(\theta_{m})-\overline{c}\right)^2}},
	\label{eq:gamma}
\end{equation}
where $C_{0}(\theta)$ is the correlation function for a massless GW background, and
$C_{\rm m}(\theta)$ is the correlation function for a massive GW background, 
which maximizes the detection significance $S$ among possible theoretical 
correlation functions for all possible values of $m_{\rm g}$. One can show that 
the optimal statistical decision rule is \citep{Kassam88}, 

\begin{displaymath}
	\left\{\begin{array}{c}
	\textrm{Choose H: if } \gamma\le\gamma_{\rm th}, \\
	\textrm{Choose F: if } \gamma < \gamma_{\rm th},
\end{array} \right.
\end{displaymath}
where the $\gamma_{\rm th}$ is the threshold of statistical decision, which is 
determined to guarantee the false alarm rate constraint $P_{\rm f}\le P_{\rm 
th}=0.1\%$.

We determine the threshold $\gamma_{\rm th}$ by a third set of
Monte-Carlo simulations. These simulations take following steps.
First, a \emph{massless} GW background is generated, then we search
for the matched value of $m_{\rm g}$, such that the corresponding
correlation function $C_{\rm m}(\theta)$ maximizes the detection
significance $S$. Then we use this $C_{\rm m}(\theta)$ to calculate
the statistic $\gamma$. We repeat this, recording the values of
$\gamma$, to establish the statistical distribution of $\gamma$
values. We take the value of the threshold $\gamma_{\rm th}$ to be
that for which there is less than a $0.1\%$ chance of getting
$\gamma>\gamma_{\rm th}$ for the case of $m_{\rm g}=0$.

After we determined the $\gamma_{\rm th}$, a fourth Monte-Carlo
simulation is used to calculated $P_{\rm d}$. This simulation is very
similar to the previous one, except that a \emph{massive} GW
background is generated. We repeat the simulation $10^{3}$ times
and take as $P_{\rm d}$ the probability of getting 
$\gamma>\gamma_{\rm th}$.

\begin{figure*}[ht]
	\centering
		\includegraphics[totalheight=5in]{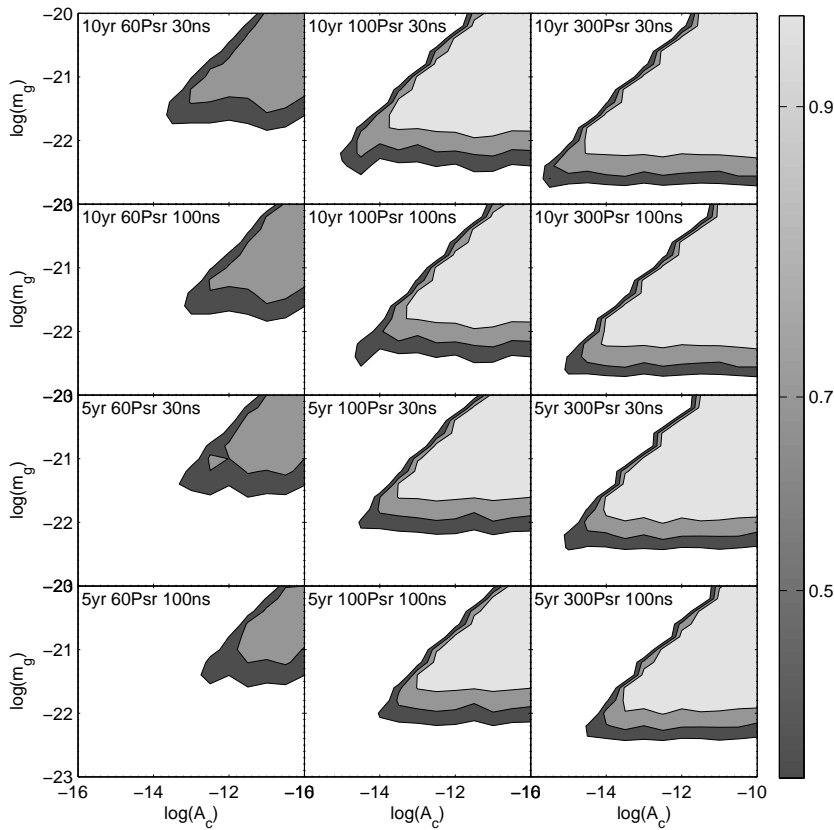}
	\caption{Contours of 50\%, 70 \% and 90\% detection rates for 
	graviton mass. The white areas are the parameter space for less than 
	50\% detection rate; the light gray regions are the parameter space with 
	 more than 90\% chance of detecting graviton mass. The false alarm rate 
	is fixed at 0.1\% for all calculations. The time span between two 
	successive observation is two weeks. The horizontal axis is the base 10 
	logarithm of the characteristic strain for the GW background. The vertical 
	axis is the logarithm of graviton mass in unit of eV. The number of pulsars
	in the timing array, the total time span of observation, and the level of 
	intrinsic pulsar timing noise are given in the legend of each panel. For 
	all the results, we use $\alpha=-2/3$.}
	\label{fig:decprob}
\end{figure*}

We summarize the results of the simulation in Fig.~\ref{fig:decprob},
which are gray-scale contour plots for the detection rate $P_{\rm d}$
for different scenarios of using 60, 100, and 300 pulsars
respectively. The corresponding parameters are given in the legend of
each panel. Intuitively, the necessary conditions for a positive
detection of a graviton mass should be first, that the GW is strong
enough for the GW to be detected, and second, that the graviton mass
is large enough to change the shape of correlation function. This
intuition is confirmed by our simulations, which show that the high
detection rate concentrates in the upper right corner of each panel,
where both graviton mass and GW amplitude are large enough.

From Fig.~\ref{fig:decprob} we can also see that we need at least 60 pulsars to 
be able to tell the difference between a massive GW background and a massless 
one.
For 5-year observations of 100 pulsars we can start to detect a graviton 
heavier than $2.5\times10^{-22}$ eV and we can achieve a limit of $m_{\rm 
g}=10^{-22}$ eV by using 5-year observations of 300 pulsars. We can achieve 
levels of $10^{-22}$ eV and $5\times 10^{-23}$ eV in 10-year observations using 
100 and 300 puslars respectively.

We also note that there is a positive correlation between the minimal detectable 
graviton mass and the GW background amplitude, as shown by the leftwards lead 
edge of the contours, in other words, we need a larger GW amplitude such that 
the GW background can be detected, if gravitons are more massive. As 
discussed above, this correlation is due to the reduction of pulsar timing 
residuals due to massive graviton.

\section{Conclusion and Discussion}
\label{sec:conc}

Current efforts to detect gravitational waves using radio pulsar
observations make use of correlated arrival time fluctuations
induced by the presence of a stochastic GW background. Einstein's
general theory of relativity makes a specific prediction for the
angular dependence of the correlation. The power spectrum of the
induced fluctuations is given by the power spectrum of the
gravitational wave background, determined by the physical processes of
the generation process, divided by the square of the GW frequency. In this 
paper, we showed that the form of the expected
correlation function, as well as the power spectrum of the induced
timing fluctuations, will be significantly altered if the graviton had a
non-zero mass. Only the transverse traceless GW modes
were considered in this analysis. 

A non-zero graviton mass introduces a cut-off frequency below which no
GWs will propagate. If we consider a fixed amount of time over which
the correlation will be measured, five years for example, a small
increase in the graviton mass will actually increase our ability to
detect a strong background. Here, we are assuming that the background is
being generated by an ensemble of supermassive black hole
binaries. This effect is due to the fact that the presence of the
graviton mass will act to flatten out the spectrum of the induced time
residuals (see Eq.~(\ref{srfg}) and Fig.~\ref{fig:eta}), thus making
the strong background easier to detect. As the mass is increased, the cut-off
frequency will increase. Once this cut-off frequency is above the
inverse of the observing time, the sensitivity starts to fall of
dramatically. This is due to the fact that pulsar timing is most
sensitive to the lowest observable frequencies and the presence of a
massive graviton removes all gravitational wave power at frequencies
below the cut-off frequency.

A non-zero graviton mass will also change the shape of the angular
dependent correlation function. In order to understand this, consider two
GW waves traveling near the direction of the line of sight between the
Earth and a given pulsar. One GW is traveling towards the pulsar, and the
other is traveling towards the Earth. The induced timing fluctuations will
be very different for each of these waves. The GW traveling towards the
Earth will induce higher amplitude fluctuations then its counter part
traveling towards the pulsar. Now, consider the same case but with a
massive graviton. As the GW frequency approaches the cut-off frequency,
the GW wavelength get arbitrarily large.  Assume that the GW wavelength
is much larger than the Earth-pulsar distance. In this case, both GWs look
exactly the same as far as the Earth-pulsar detector is concerned. Hence,
both waves induce the same timing fluctuations. This symmetry between
the two GW directions implies that the timing fluctuations from two
pulsars near each other on the sky or $180^\circ$ apart will have the
same amount of correlation, unlike the massless graviton case. Thus,
as the graviton mass increases, the correlation curve will become more
symmetric about $90^\circ$.

Since the graviton mass changes the expected correlation function, we
can detect the massive graviton by measuring the shape of correlation
function.{ The optimal detection algorithm differentiating between a massive 
and  a massless GW background was constructed in this paper.} Using this 
algorithm, we found
that at least 60 pulsars are required to discriminate between a GW
background made up of massive and massless gravitons. Note that,
like the case of discriminating various polarization modes of GWs
\citep{LJR08}, the minimum number of pulsars is insensitive to the
intrinsic timing noise level or the total observing time.

{ Here, we are mainly focusing on answering following question:
What is the threshold of graviton mass, such that the detector, we constructed 
in the paper, will find out that the GW background is composed of massive 
gravitons rather than massless gravitons? We answered this statistical detection 
question in this paper. The related parameter estimation question \footnote{How 
well can we measure the graviton mass or put uplimit on the graviton mass?}  
will be investigated in the future works.  }

In this paper, we treat the graviton mass in the sense of a GW dispersion
relation. A better graviton mass upper limit ($m_{\rm g}\le 2\times
10^{-26}$ eV) can be achieved by observing the GW dispersion using
Laser Interferometer Space Antenna (LISA) \citep{SW09}. Besides GW 
dispersion based techniques, there are other good upper limits ($2\times 
10^{-29} h_{0}$ eV) from galaxy
cluster observations \citep{GN74}. However since some gravity
theories \citep{Maggiore08} contain mass terms for $h_{ab}$, but maintain the 
scalar sectors, the $m_{\rm g}$ upper limits from GW dispersion observations
are independent of the upper limits from Yukawa potential
experiments such as the Solar System and the galaxy cluster observations.

In summary, for the task of detecting the massive graviton using a pulsar
timing array, there is one critical requirement: a large sample of
stable pulsars. Thus the on-going and upcoming projects like the Parkes
Pulsar Timing Array \citep{HBBC09}, European Pulsar Timing Array
\citep{SKLDJ06}, NANOGrav \citep{JEN09}, the Large European
Array for Pulsars \citep{BVK09}, the Five-hundred-meter Aperture
Spherical Radio Telescope \citep{NWZZJG04, SLKMSJN09} and the Square
Kilometer Array (SKA) will offer unique opportunities to find, to time these 
pulsars; and to detect the GW background and measure its properties.

\section{Acknowledgment}
K. J. Lee gratefully acknowledge support from ERC Advanced Grant ``LEAP``, Grant 
Agreement Number 227947 (PI M. Kramer), F. A. Jenet and R. H. Price gratefully acknowledge 
support from the NSF under grants AST0545837
and PHY0554367, and support from the UTB Center for
Gravitational Wave Astronomy. We also thank M. Pshirkov and N. Straumann for 
very useful comments.

\clearpage
\appendix

\section{Power spectra of timing residual}
\label{app:sing}
The power spectra of timing residuals is calculated by integrating 
Eq.~(\ref{eq:powr}). Here we give the details of the calculation.

\begin{figure}[ht]
	\centering
		\includegraphics[totalheight=2.5in]{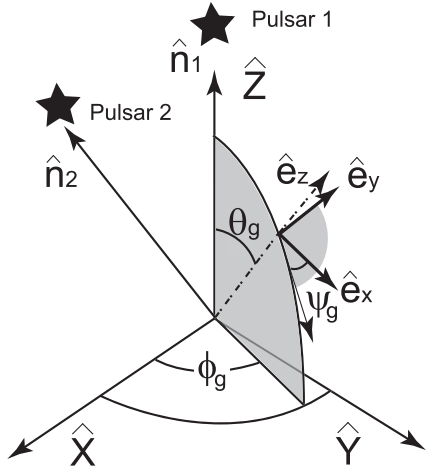}
	\caption{The geometric configuration of the coordinates and unit vectors 
	used here. The $\hat{X}$,~$\hat{Y}$~and~$\hat{Z}$ are the coordinate unit 
	vectors, $\hO$ is the propagation direction of the GW, and
	$\hp_{1}$ and $\hp_{2}$ are unit vectors pointing to the pulsars.}
	\label{figgeoconf}
\end{figure}

In Fig.~\ref{figgeoconf}, the
components of $\hO$ in the $\hat{\bf X}^a=(\hat{\bf X},\hat{\bf
 Y},\hat{\bf Z})$ frame can be seen to be
$(\sin\theta_{g}\cos\phi_{g},$
$\sin\theta_{g}\sin\phi_{g},\cos\theta_{g})$, where the
$\theta_{g}$ and $\phi_{g}$ are respectively the polar angle and
azimuthal angle of the GW propagation vector. To proceed, we 
need the components of the polarization tensors in the $\hat{\bf X}^a$
frame. The transformation from the components $\peps_{0,cd}$ given in
the GW frame $\hat{\bf e}_b =(\hat{\bf e}_x ,\hat{\bf e}_y,\hat{\bf
 e}_z)$ of Eq.~(\ref{eq:polten}) is made with $\peps^{P}_{ab}
=T_{ca}T_{db}\,\peps^{P}_{0,cd}$, where $T_{ca}=\hat{\bf
 e}_c\cdot\hat{\bf X}_a$ has components
\begin{equation}\label{eq:xformmatrix}
\left({
\begin{array}{c c c}
\cos\theta_{g}\cos\phi_{g}\cos{\psi_{g}}-\sin\phi_{g}\sin{\psi_{g}} 
&\cos\theta_{g}\cos\psi_{g}\sin\phi_g+\cos\phi_g\sin\psi_g
&-\cos\psi_{g}\sin\theta_{g} \\
-\cos\psi_{g}\sin\phi_{g}-\cos\theta_{g}\cos\phi_{g}\sin\psi_{g} 
&\cos\phi_g\cos\psi_g-\cos\theta_{g}\sin\phi_{g}\sin\psi_g & 
\sin\theta_g\sin\psi_g \\
\cos\phi_{g}\sin\theta_{g} &\sin\theta_g\sin\phi_g &\cos\theta_{g}
\end{array}
}\right)\,.
\end{equation}

Since the GW background is isotropic, with no loss of
generality one can choose $\hp= \{0, 0, 1\}$ so that
\begin{equation}
	\wg+c\hp\cdot {\rm \bf \kg} = (1+\zeta \cos \theta_{\rm g})\,\wg,
\end{equation}
where 
\begin{equation}
 \label{eq:zetadef}
 \zeta =\sqrt{1-\frac{\mg^2 c^4}{\wg^2 \hbar^2}\;}=\sqrt{1-\frac{f_{\rm 
 cut}^2}{f_{\rm g}^2}\;}\,,
\end{equation}
for $f_{\rm g}\geq f_{\rm cut}$\,.
The term $\peps ^{P}_{ij}\hp^i \hp^j$ now simplifies to
\begin{eqnarray}
	\peps ^{+}_{ij}\hp^i \hp^j &=&\sin^2\theta_{\rm g} \cos(2\psi_{\rm g}) 
	\nonumber \\
	\peps ^{\times}_{ij}\hp^i \hp^j &=&-\sin^2\theta_{\rm g} \sin(2\psi_{\rm g})\,. 
\end{eqnarray}
After the ensemble average over the polarization angle $\psi_{\rm g}$, the 
integration of
Eq.~(\ref{eq:powr}) gives
\begin{equation}
  S_{\rm R}(f_{\rm g})=\sum_{P=+,\times}\frac{|h_{c}^{P}(f_{\rm g})|^2}{128 
  \pi^3 f_{\rm g}^3} \int_{0}^{\pi} d\theta_{\rm g} \int_{0}^{2\pi} d\phi_{\rm 
  g} \frac{\sin^5\theta_{\rm g}}{(1+\zeta \cos\theta_{\rm g})^2} 
  \left[1-\cos\left(\frac{D\wg}{c} (1+\zeta \cos\theta_{\rm g})\right)\right]
\end{equation}

For practical pulsar timing array, we have
$\Phi_0=D\wg/c\gg 1$, i.e.\,, the GW wave length is much smaller than the pulsar-earth 
distance, so one gets \begin{equation}
  \int_{-1}^{1} \frac{(1-\mu^2)^2}{(1+\zeta \mu)^2}\left[1-\cos \left( \Phi_0 
  (1+\zeta \mu)\right)\right] d\mu=
\frac{4 \left[-4 \zeta ^3+6 \zeta +3 \left(\zeta ^2-1\right) \log 
\left(\frac{\zeta +1}{1-\zeta }\right)\right]}{3
\zeta ^5} + O\left(\Phi_0^{-2}\right).
\end{equation}
Finally, the power spectra of the pulsar timing residuals then becomes
\begin{equation}
  S_{\rm R}(f_{\rm g})=\sum_{P=+,\times}\frac{|h_{c}^{P}(f_{\rm g})|^2}{24 \pi^2 
  f_{\rm g}^3} \eta(f_{\rm g})\,.
	\label{eq:sr}
\end{equation}
For $f_{\rm g}>f_{\rm cut}$, $\eta$ is given by
\begin{equation}
  \eta(f_{\rm g})=\frac{-4 \zeta ^3+6 \zeta +3 \left(\zeta ^2-1\right) \log 
  \left(\frac{\zeta +1}{1-\zeta }\right)}{2 \zeta^5}\,,
\end{equation}
with $\zeta$ is given by Eq.~(\ref{eq:zetadef})\,. For $f_{\rm g}\leq f_{\rm 
cut}$ the gravitational waves cannot propagate, and we take $\eta=0$.

\section{Angular Dependent Correlation Function for Pulsar Timing Residuals}
\label{app:hel}

The correlation function $C(\theta)$ between pulsar `$a$' and `$b$' defined in 
this paper is
\begin{equation}
	C(\theta)=\frac{\langle R_{a}(t) R_{b}(t)\rangle}{\sigma_{a} \sigma_{b}},
\label{eq:cordef}
\end{equation}
where $\theta$ is the angle between the direction to pulsar `$a$' and the 
direction to pulsar `$b$'. The $R_{a}(t)$ and $R_{b}(t)$ are the GW induced 
timing residuals for pulsar `$a$' and `$b$' respectively. The $\sigma_{a}$ and 
$\sigma_{b}$ are the RMS value for the timing residual $R_{a}(t)$ and 
$R_{b}(t)$. One
has (see \cite{LJR08} for a similar calculation)

\begin{equation}
	C(\theta)=\frac{1}{\sigma_{a} \sigma_{b}} \sum_{P=+,\times} \left\langle 
	\int_{0}^{\infty} \frac{|h_{c}^{P}(f_{\rm g})|^2}{64 \pi f_{\rm g}} df_{\rm 
	g} \int d\Omega \frac{\peps ^{P}_{ij}\hp_{a}^i \hp_{a}^j }{\wg+c {\rm \bf 
	\kg \cdot \hp_{a}}} \frac{\peps ^{P}_{ij}\hp_{b}^i \hp_{b}^j }{\wg+c {\rm 
	\bf \kg \cdot \hp_{b}}} {\cal P}_{ab} \right\rangle,
	\label{eq:corful}
\end{equation}
where ${\cal P}_{ab}$ is given by
\begin{eqnarray}
 {\cal P}_{ab}&=& 1- \cos\Phi_{a}-\cos\Phi_{b}+\cos(\Phi_{a}-\Phi_{b}); 
 \nonumber \\
  \Phi_{a}&=&\frac{D_{a}}{c} (\wg+c\kg\cdot \hp_{a}); \nonumber \\
  \Phi_{b}&=&\frac{D_{b}}{c} (\wg+c\kg\cdot \hp_{b}).
	\label{eq:ptdef}
\end{eqnarray}

Equation~(\ref{eq:corful}) shows that the correlation function is composed of 
two major integrations, one over the GW frequency $df_{\rm g}$, and the one over 
solid angle $d\Omega$.  For the massless graviton case, these two parts are not 
mixed, since $\kg$ is linear in $\wg$. But $\kg$ is non-linear in $\wg$ for the 
case of a massive graviton.  The nonlinear dependence in Eq.~(\ref{eq:dispf}) 
mixes the spatial integral with the frequency one. Thus the shape of $C(\theta)$ 
depends on both the graviton mass and the frequency spectra of GWs, which also 
depends on the observation schedule. Thus, it is unlikely that one can integrate 
Eq.~(\ref{eq:corful}) analytically. 

\end{document}